\documentclass{IEEEtran4PSCC}

\ifCLASSINFOpdf
   \usepackage[pdftex]{graphicx}
\else
   \usepackage[dvips]{graphicx}
\fi

\usepackage[cmex10]{amsmath}
\usepackage{amsfonts, amssymb}
\usepackage{cellspace}
\usepackage[table]{xcolor}
\usepackage{makecell}
\setcellgapes{3.5pt}
\usepackage{subcaption}
\usepackage{xcolor}
\usepackage{booktabs}
\usepackage{multirow}
\usepackage{makecell}
\usepackage{soul}

\usepackage{algorithm}      %
\usepackage{algpseudocode}  
\newcommand{\hide}[1]{}

\newcommand{\bit}{\begin{compactitem}}
\newcommand{\eit}{\end{compactitem}}
\newcommand{\ben}{\begin{compactenum}}
\newcommand{\een}{\end{compactenum}}

\usepackage{xcolor}

\newcommand{\norm}[1]{\lVert #1 \rVert}

\newtheorem{theorem}{Theorem}
\makeatletter
\let\old@ps@headings\ps@headings
\let\old@ps@IEEEtitlepagestyle\ps@IEEEtitlepagestyle
\def\psccfooter#1{%
    \def\ps@headings{%
        \old@ps@headings%
        \def\@oddfoot{\strut\hfill#1\hfill\strut}%
        \def\@evenfoot{\strut\hfill#1\hfill\strut}%
    }%
    \def\ps@IEEEtitlepagestyle{%
        \old@ps@IEEEtitlepagestyle%
        \def\@oddfoot{\strut\hfill#1\hfill\strut}%
        \def\@evenfoot{\strut\hfill#1\hfill\strut}%
    }%
    \ps@headings%
}
\makeatother

\psccfooter{%
        \parbox{\textwidth}{\hrulefill \\ \small{24th Power Systems Computation Conference} \hfill \begin{minipage}{0.2\textwidth}\centering \vspace*{4pt} \includegraphics[scale=0.06]{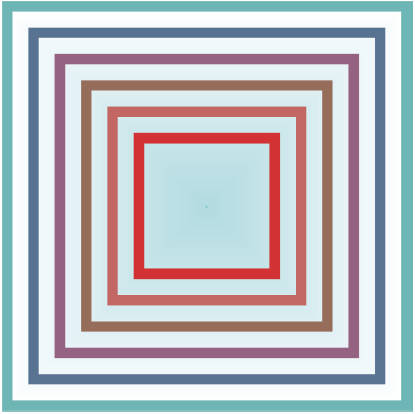}\\\small{PSCC 2026} \end{minipage} \hfill \small{Limassol, Cyprus --- June 8 -- June 12, 2026}}%
}

\begin{document}
%
\title{
Optimal Bidding and Coordinated Dispatch of Hybrid Energy Systems in Regulation Markets}

\author{
\IEEEauthorblockN{Tanmay Mishra, Dakota Hamilton, Mads R. Almassalkhi}
\IEEEauthorblockA{Department of Electrical and Biomedical Engineering, University of Vermont, USA\\
\{tmishra, dhamilt6, malmassa\}@uvm.edu}
}


\maketitle

\begin{abstract}
The increasing integration of renewable energy sources and distributed energy resources (DER) into modern power systems introduces significant uncertainty, posing challenges for maintaining grid flexibility and reliability. Hybrid energy systems (HES), composed of controllable generators, flexible loads, and battery storage, offer a decentralized solution to enhance flexibility compared to single centralized resources. This paper presents a two-level framework to enable HES participation in frequency regulation markets. The upper level performs a chance-constrained optimization to choose capacity bids based on historical regulation signals. At the lower level, a real-time control strategy disaggregates the regulation power among the constituent resources. 
This real-time control strategy is then benchmarked against an offline optimal dispatch  to evaluate flexibility performance. 
Additionally, the framework evaluates the profitability of overbidding strategies and identifies thresholds beyond which performance degradation may lead to market penalties or disqualification. The proposed framework also compare the impact of imbalance of power capacities on performance and battery state of charge (SoC) through asymmetric HES configurations.
\end{abstract}

\begin{IEEEkeywords}
flexibility, hybrid energy systems, controllable load, battery storage, frequency regulation, optimal bidding, energy market.
\end{IEEEkeywords}
\thanksto{\noindent This material is based upon work supported by the U.S. Department of Energy's Office of Energy Efficiency and Renewable Energy (EERE) under the Solar Energy Technologies Office Award Number DE-EE0010147. The views expressed herein do not necessarily represent the views of the U.S.Department of Energy or the United States Government.}
\vspace{-0.5cm}

\section{Introduction}
The modern power system is undergoing rapid transformation driven by the integration of variable renewable energy sources (RES) such as solar and wind, coupled with the proliferation of distributed energy resources (DER) including battery storage, electric vehicles (EV), and flexible controllable loads~\cite{Yang2022}. This transition poses significant challenges for maintaining grid stability and reliability due to the variability and uncertainty of RES, along with bi-directional power flows and shifting load profiles introduced by widespread DER integration~\cite{Liang2016,Hou2020}.
These evolving dynamics highlight the need to revisit the concept of system flexibility centering around these resources, as traditional measures may no longer capture the diverse and rapid response capabilities required for modern power systems \cite{Walied2019}. 

\begin{figure}[t]
\centering
\includegraphics[width=1.0\linewidth]{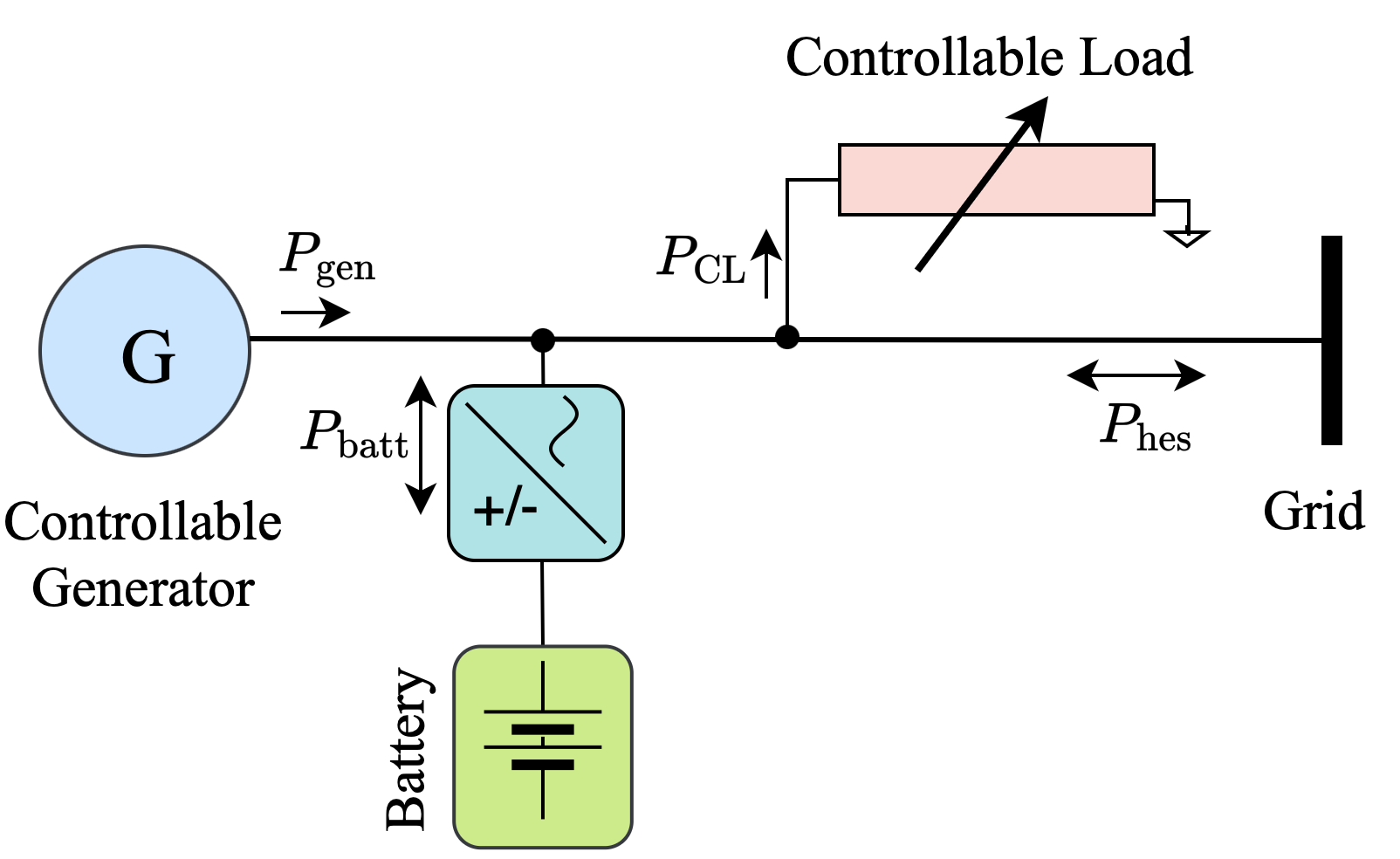}
\caption{Hybrid energy system consisting of a controllable generator, battery storage, and a controllable load connected to an utility grid.}
\label{fig:HES}
\centering
\end{figure}

Flexibility in this context can be defined as the ability to export and import a required power (e.g, provide $\pm M$ MW for frequency regulation for a given duration of time), quantified by the feasible set of maximum power and net energy within a specified dispatch window~\cite{Nosair2015,Hadi2022}. While large scale battery storage has often been the go-to solution, recent grid events (e.g., California's ``duck curve" over-generation~\cite{Calero2022}) have exposed the limitations of relying solely on centralized, single-resource flexibility. These events underscore the high costs, under-utilization, and inability of such singular solutions to address the location-specific and temporally nuanced needs of a dynamic grid \cite{Mohandes2019}. 
This highlights a critical need for a multi-resource system such as hybrid energy systems (HES), as shown in Fig.~\ref{fig:HES}, which coordinates a controllable generator and loads along with energy storage to provide flexibility for diverse grid services \cite{Cheng2023, Murphy2021}. 

However, this potential solution introduces new challenges, such as: \textit{How can the flexibility of an HES be maximized, and how can its diverse resources be coordinated in real-time to effectively participate in frequency regulation markets?} In practice, maximizing flexibility is closely tied to determining the optimal dispatch capacity, large enough to capture market value but constrained by the minimum performance requirement set by the market regulators~\cite{Huang2025,Lyu2023}. Achieving this balance requires an HES to operate with a real-time dispatch strategy that complements operational limits of its assets while collectively maintaining the desired performance score~\cite{Guannan2017}. 
This interdependence of choosing the optimal bid and dispatch becomes even more challenging when the regulation signal is not known in advance.
This motivates the need for bidding strategies that are robust to regulation signal uncertainty while ensuring market compliance~\cite{Bolun2018}. 
One approach is to treat these two (optimal bidding and dispatch) as separate problems~\cite{Behzad2019}, or to assume that resources can always dispatch the optimal bid obtained through scenario-based techniques without violating their operational limits~\cite{Sen2019}; however, such simplifications often lead to suboptimal bids that either violate performance requirements or fail to fully utilize the system’s flexibility. 

To overcome these limitations, a framework that jointly considers performance requirements, asset constraints, and market participation objectives is required. One promising direction is a bi-level formulation in which bidding and dispatch are inherently coupled~\cite{Bolun2018}. In this approach, the capacity bidding problem is solved using chance constraints to a guaranteed probability of success for a minimum performance score, while a real-time control strategy determines the optimal power dispatch during market operation. In~\cite{Bolun2018}, this framework is demonstrated for a battery participation in frequency regulation market where battery aging is explicitly considered as a key factor influencing the bidding and dispatch decisions.
Building on this foundation, the main contributions of the paper are as follows:
\begin{itemize}
    \item[1)] A chance-constrained bidding strategy to maximize HES participation in frequency regulation markets.
    \item[2)] A real-time control strategy is proposed to disaggregate the regulation signal among controllable generators, flexible loads, and battery storage in an optimal manner.
    \item [3)]Simulation-based analysis benchmarks the proposed real-time control strategy against the offline optimal dispatch and evaluates tradeoffs in profitability and performance across different bidding scenarios. The study also quantifies how symmetric and asymmetric HES configurations influence achievable performance and SoC sustainability.
    
    
    
\end{itemize}

The rest of this paper is organized as follows: Section~\ref{Component modeling} presents the modeling of HES components. Section~\ref{Market setting} describes the pay-for-performance market framework. Section~\ref{Bid_and_control} outlines the proposed optimal bidding and real-time control strategy for power allocation among HES assets. Section~\ref{case_studies} evaluates performance through case studies. Finally, Section~\ref{Conclusion} summarizes key findings and discusses future work.

\section{HES Component Modeling}
\label{Component modeling}

An HES combines multiple controllable resources, such as generators, storage, and flexible loads, to provide real-time grid flexibility beyond what a single resource can offer. 
We model the HES shown in Fig.~\ref{fig:HES}, consisting of three key components: a controllable generator, an energy storage device (battery), and a controllable load. Their power variables and operational constraints are defined below.

\subsection{Controllable generator}
The generator may represent an intermittent renewable source (e.g., solar or wind) or a dispatchable unit (e.g., natural gas or diesel genset). 
The output power of the generator at each discrete time step $k$ is denoted by $P_{\text{gen}}[k]$, and is constrained by
\begin{equation}
    \underline{P}_{\text{gen}} \leq P_{\text{gen}}[k] \leq \overline{P}_{\!\text{gen}}\,,
\end{equation}
where $\underline{P}_{\text{gen}}$ and $\overline{P}_{\!\text{gen}}$ are the minimum and maximum generator power outputs.
For intermittent sources such as solar photovoltaics (PV) or wind, the power limit is time-varying, can be denoted as $\overline{P}_{\!\text{gen}}[k]$  however, in this work, a constant power limit is assumed for simplicity.

\subsection{Battery energy storage}
The battery dispatch capability is defined using power and energy bounds with total battery dispatch power $P_{\text{batt}}[k] = P^{\text{d}}_{\text{batt}}[k] +P^{\text{c}}_{\text{batt}}[k]$, where $P^{\text{d}}_{\text{batt}}[k] \ge 0$ and $P^{\text{c}}_{\text{batt}}[k] \le 0$  represent discharging and charging powers, respectively. 
The instantaneous battery output power is constrained by
\begin{equation}
    -\overline{P}_{\text{batt}} \leq P_{\text{batt}}[k] \leq \overline{P}_{\text{batt}}\,,
\end{equation}
where $\overline{P}_{\text{batt}}$ is the rated power capacity of the battery.
 The battery state of charge (SoC), denoted by $E[k]$, dynamics are given by \cite{Arash2025}:
\begin{equation}
    E[k+1] = E[k] 
        - \left(\eta_{\text{c}} P_{\text{batt}}^\text{c}[k]  
        + \frac{1}{\eta_{\text{d}}} P_{\text{batt}}^\text{d}[k]\right) \frac{\Delta t}{\overline{P}_{\text{batt}}}\,,
\end{equation}
where $\Delta t$ is the size of the discrete time step, and $\eta_{\text{c}}$ and $\eta_{\text{d}}$ denote charging and discharging efficiencies, respectively.
The battery is also subject to SoC limits:
\begin{equation}
    \underline{E} \leq E[k+1] \leq \overline{E}\,.
\end{equation}

\subsection{Controllable load}
The controllable load may represent data center, process-driven loads such as an electrolyzer, or EV charging station. The load power consumption, denoted by $P_{\text{CL}}[k]$, is subject to
\begin{equation}
    0 \leq P_{\text{CL}}[k] \leq \overline{P}_{\text{CL}}\,.
\end{equation}

\subsection{Total HES power output}

By integrating these components under a coordinated control structure, an HES capable of flexible grid interaction is modeled.
Thus, the total HES power output $P_\text{hes}[k]$ is
\begin{equation}
    P_{\text{hes}}[k] = P_{\text{gen}}[k] - P_{\text{CL}}[k] + P_{\text{batt}}^\text{d}[k] + P_{\text{batt}}^\text{c}[k]\,.
\end{equation}


\section{Regulation Market Setting}
\label{Market setting}

We consider HES participation in a pay-for-performance regulation market (cleared hourly), where the HES operator submits a capacity bid $C$ (in MW) ahead of the operating hour.
If the HES is cleared to participate, it is asked to follow a normalized regulation signal $r[k] \in [-1,1]$, scaled by the capacity bid $C$ to determine the actual power reference, for all discrete time steps $k \in \mathcal{K} = \{0,1,\dots,N-1\}$ of the operational window. 
That is, the HES must adjust its power dispatch $P_\text{hes}[k]$ (in MW) to follow the signal $Cr[k]$. 
At the end of the operational period, the tracking accuracy of the HES is evaluated using the performance score $x_{\text{p}} \in [0,1]$, as defined in~\cite{Bolun2018}:
\begin{equation}
\label{eq:perf_score}
x_{\text{p}} = 1 - \frac{\norm{C \mathbf{r} - \mathbf{p}_\text{hes}}_1}
              {C \norm{\mathbf{r}}_1}\,,
\end{equation}
where vectors $\mathbf{r}$ and $\mathbf{p}_\text{hes}$ are defined as
\begin{gather}
    \mathbf{r}=\left[r[0],\dots,r[N-1]\right]^\top\,,\\
    \mathbf{p}_\text{hes}=\left[P_\text{hes}[0],\dots,P_\text{hes}[N-1]\right]^\top\,.
\end{gather}
The HES market participant is then compensated based on the market clearing price $\lambda_{\text{c}}$, their capacity bid $C$, performance score $x_\text{p}$, the mileage of the regulation signal $M$, and the mileage clearing price $\lambda_{\text{m}}$~\cite{Bolun2016}. 
Therefore, we model the revenue $R$ (in~\$) received by the HES market participant as
\begin{equation}
\label{eq:revenue}
    R = C x_{\text{p}}(\lambda_{\text{c}} +\lambda_{\text{m}} M)\,,
\end{equation}
where the mileage $M$, is defined as
\begin{equation}
M = \sum_{k=0}^{N-2} \left| r[k+1] - r[k] \right|\,.
\end{equation}

Additionally, poor performance below a minimum threshold, denoted by $\underline{x}_\text{p}$, can result in severe financial penalties, such as loss of compensation or market disqualification.
Since $C$ directly affects $x_\text{p}$, the participant must choose the largest bid that can be sustained without violating the performance threshold. 
If the regulation signal $\mathbf{r}$ and prices ($\lambda_\text{c}$ and $\lambda_\text{m}$) were deterministic or known in advance, this would reduce to a straightforward bid optimization problem: 
\begin{subequations}
\begin{align}
    \max_{C}~~& C x_{\text{p}}(\lambda_{\text{c}} + \lambda_{\text{m}} M)\,, \label{eq:obj_func0}\\
    \text{s.t.}~~& x_\text{p} \ge \underline{x}_\text{p}\,, \\
    & x_\text{p} = 1 - \frac{\norm{C \mathbf{r} - \mathbf{p}^{*}_\text{hes}(C,\mathbf{r})}_1}
              {C \norm{\mathbf{r}}_1}\,, \\
    & \mathbf{p}^{*}_\text{hes}(C,\mathbf{r}) = \operatorname{arg}\,\min_{\mathbf{g}\in\mathcal{G}}\, \norm{C \mathbf{r} - \mathbf{g}(C,\mathbf{r})}_1\,, \label{eq:offline_opt_control}\\
    & 0 \le C \le C_\text{max}\,,
\end{align}
\end{subequations}
where $C_\text{max}$ is the maximum allowable bid capacity, determined by market-specific policies.
Here, $\mathbf{p}^{*}_\text{hes}(C,\mathbf{r})$ denotes the optimal HES dispatch strategy which minimizes tracking error while respecting the operational constraints of HES components.
The set $\mathcal{G}$ denotes the set of feasible dispatch strategies (see Sec.~\ref{sec:offline_control} for details).

In practice, the capacity bid $C$ is committed before the operating hour, while the performance score $x_\text{p}$ is realized afterward \cite{Bolun2018}.
Since the regulation signal $\mathbf{r}$ is uncertain at the time of bidding, the resulting performance score $x_\text{p}$  is also inherently uncertain. 
Thus, in the subsequent section, we discuss the optimal bidding and real-time dispatch strategies under this uncertainty and benchmark these strategies against an offline optimal dispatch strategy.

\section{Optimal Capacity Bidding and Dispatch Strategy}\label{Bid_and_control}

Since the regulation signal $\mathbf{r}$ is not known in advance, determining the optimal capacity to bid and the optimal HES dispatch is challenging. 
Overestimating $C$ can cause persistent tracking errors, reducing the performance score $x_\text{p}$ and potentially leading to financial penalties or market disqualification. Underestimating $C$ results in conservative operation and lost revenue. 
Therefore, the optimal bidding strategy must balance maximizing expected revenue with maintaining performance compliance under uncertainty. 
Furthermore, designing a real-time control strategy to allocate power among individual HES assets adds another layer of complexity. 
To address these challenges, we formulate a bi-level chance-constrained optimization problem: the outer level formulation addresses the bidding problem, and is solved by leveraging historical variations in $\mathbf{r}$; the inner level implements a real-time control strategy that, for a given capacity bid $C$, allocates power among the HES components to ensure optimal tracking of the regulation signal while maintaining feasible operation.

\subsection{Chance-constrained problem formulation}

To capture the trade-off between maximizing revenue and maintaining performance above a minimum threshold, we model the bidding decision as a chance-constrained optimization problem, which ensures that the HES can meet reliability requirements under uncertainty in the regulation signal while selecting the largest feasible capacity bid. The mathematical formulation is given as,
\begin{subequations}
\label{eq:cc_problem}
\begin{align}
    \max_{C,\,g\in\mathcal{G}}~~&C\,\mathbb{E}\left[ x_\text{p} \right]\,, \label{eq:cc_obj}\\
    \text{s.t.}~~&\mathbb{P}(x_\text{p} \ge \underline{x}_\text{p}) \ge \gamma\,, \label{eq:chance_constraint}\\
    &x_\text{p} = 1 - \frac{\norm{C \mathbf{r} - \mathbf{p}_\text{hes}(C,\mathbf{r})}_1}
              {C \norm{\mathbf{r}}_1}\,, \\
    &P_\text{hes}[k] = g(C,r[k])\,, \quad \forall k \in \mathcal{K}\,,\label{eq:cc_dispatch} \\
    &0 \le C \le C_\text{max}\,.
\end{align}
\label{Bid_optimization}
\end{subequations}
The overall objective is to maximize expected market revenue~\eqref{eq:revenue}; however, the prices ($\lambda_\text{c}$ and $\lambda_\text{m}$) and the mileage $M$ are unknown at the time of bidding.
Furthermore, we assume these quantities are not affected by the capacity bid or HES dispatch strategy.\footnote{Note that the mileage $M$ only depends on the regulation signal. For the prices, we assume that the HES market participant behaves as a price-taker (i.e., it offers a zero price to ensure it is cleared and accepts any clearing and mileage prices).}
Therefore, we focus on maximizing the expected value of that revenue component which we can control (i.e., $Cx_\text{p}$), resulting in the objective function~\eqref{eq:cc_obj}.

The chance constraint~\eqref{eq:chance_constraint} ensures that the HES achieves a performance score above  $\underline{x}_\text{p}$ with confidence $\gamma$. This probabilistic guarantee provides robustness against variability in the regulation signal, improving the reliability of HES participation in frequency regulation markets. Finally, the constraint~\eqref{eq:cc_dispatch} requires that the HES dispatch at time~$k$ depends only on the regulation signal at time~$k$.
This is in contrast to the deterministic optimal HES dispatch~\eqref{eq:offline_opt_control}, which may utilize future information about the regulation signal.

The problem formulation closely follows the performance-based bidding framework in \cite{Bolun2018,Guannan2016}, but extends it to hybrid systems by explicitly incorporating multiple controllable resources. To solve this problem, we work backwards by first proposing a real-time dispatch strategy $g(C,r[k])$, then combine this control strategy with historical regulation signal data to characterize empirical probability density functions for the HES performance score for different capacity bids.


\subsection{Optimal HES dispatch strategy}

In this section, we discuss optimal control strategies for dispatching HES and allocating power among HES components.
Here, we consider the capacity bid $C$ to be fixed and the goal is to minimize the tracking error while satisfying operational limits.

\subsubsection{Offline optimal dispatch}
\label{sec:offline_control}
First, we consider the optimal dispatch from~\eqref{eq:offline_opt_control}, which represents the best possible dispatch assuming perfect knowledge of the regulation signal $\mathbf{r}$. 
This offline control strategy is formulated as the following constrained optimization problem:
\begin{subequations}
\label{eq:offline_opt}
\begin{align}
\min_{\mathbf{y}}~~&\sum_{k=0}^{N-1} \lvert Cr[k] - P_{\text{hes}}[k] \rvert\,,  \\
\text{s.t.}~~& P_{\text{hes}}[k] = P_{\text{gen}}[k] - P_{\text{CL}}[k] + P_{\text{batt}}^\text{d}[k] + P_{\text{batt}}^\text{c}[k]\label{hes_power}\,,\nonumber\\
&\hspace{5.6cm}\forall k \in \mathcal{K}\,,\\
& 0 \leq P_{\text{gen}}[k] \leq \overline{P}_{\!\text{gen}}\,,\quad \forall k \in \mathcal{K}\,\,,\label{eq:gen_pow_lim}\\
& 0 \leq P_{\text{CL}}[k] \leq \overline{P}_{\text{CL}}\,,\quad \forall k \in \mathcal{K}\,,\label{eq:load_pow_lim}\\
&0 \leq P^{\text{d}}_{\text{batt}}[k] \leq \overline{P}_{\text{batt}}\,,\quad \forall k \in \mathcal{K}\,, \label{eq:batt_d_lim}\\
&-\overline{P}_{\text{batt}} \leq P^{\text{c}}_{\text{batt}}[k] \leq 0\,,\quad \forall k \in \mathcal{K}\,, \label{eq:batt_c_lim}\\
& P^{\text{c}}_{\text{batt}}[k] P^{\text{d}}_{\text{batt}}[k] = 0\,,~~\forall k \in \mathcal{K}\,,\label{eq:batt_cc}\\
&E[k+1] = E[k] - \left(\eta_\text{c} P_{\text{batt}}^\text{c}[k]+\frac{1}{\eta_\text{d}} P_{\text{batt}}^\text{d}[k]\right)\frac{\Delta t}{\overline{P}_{\text{batt}}}\,,\nonumber\\
&\hspace{3.5cm}\forall k \in \mathcal{K}\,,\quad E[0] = E_{0}\,,\label{eq:batt_dyn}\\
&\underline{E} \leq E[k+1] \leq \overline{E}\,,\quad \forall k \in \mathcal{K}\,,\label{eq:batt_soc_lim}
\end{align}
\end{subequations}
where the decision variables $\mathbf{y}$ consist of $P_{\text{gen}}[k]$, $P_{\text{CL}}[k]$, $P_{\text{batt}}^\text{d}[k]$, and $P_{\text{batt}}^\text{c}[k]$, for all times $k\in\mathcal{K}$.
The constraints~\eqref{eq:gen_pow_lim}--\eqref{eq:batt_c_lim} impose power limits of each individual asset, and the complementary constraint~\eqref{eq:batt_cc} prevents simultaneous charging and discharging of the battery.
Finally, the battery SoC dynamics and SoC limits are enforced by~\eqref{eq:batt_dyn}--\eqref{eq:batt_soc_lim}, where $E_0$ denotes the initial SoC of the battery.

This optimization problem is non-convex due to the complementary constraint~\eqref{eq:batt_cc}.
However, the problem can be reformulated into an equivalent mixed-integer linear program (MILP), which can be solved using existing commercial solvers (e.g., Gurobi)~\cite{Tanmay2024,Mazen2023}.
Furthermore, for some regulation signals, the optimal dispatch determined by~\eqref{eq:offline_opt} may be non-unique. 
That is, there may be multiple ways to allocate power among the available assets which achieve the same HES output, and thus the same tracking performance.

The offline optimal dispatch~\eqref{eq:offline_opt} is not directly implementable in practice due to its dependence on perfect knowledge of future signals.
Nonetheless, the optimal offline tracking error obtained by solving this problem, which we denote as $J^{*}_\text{off}(C,\mathbf{r})$, serves as a benchmark with which we evaluate the performance of the proposed real-time control strategy.



\subsubsection{Real-time control strategy}
The proposed real-time control strategy is given by Algorithm~\ref{alg:hes_dispatch_short}.
The strategy determines the HES dispatch at each time $k\in\mathcal{K}$ and allocates power among the available resources while respecting their power and energy constraints. 
At each timestep, the allocation prioritizes controllable generation and load, using the battery only when the required export or import cannot be met otherwise. 
This approach reduces battery cycling and mitigates large SoC deviations. 
Furthermore, we show that when battery SoC limits are not binding during the operational window, then the proposed real-time control strategy achieves the same performance as the offline optimal dispatch.

\begin{algorithm}[tb]
\caption{Proposed Real-time Control Strategy $g(C,r[k])$}\label{alg:hes_dispatch_short}
\begin{algorithmic}[1]
\renewcommand{\algorithmicrequire}{\textbf{Given:}}
\Require $C$, $r[k]$, $\overline{P}_{\!\text{gen}}$, $\overline{P}_{\text{CL}}$, $\overline{P}_{\text{batt}}$, $\eta_{\text{c}}$,$\eta_{\text{d}}$, $E[k]$, $\underline{E}$, $\overline{E}$, $\Delta t$


\vspace{0.25em}
    \State $\delta_\text{c} = \min\{\frac{1}{\eta_\text{c} \Delta t}(\overline{E}-E[k])\,, 1\}$
    \vspace{0.25em}
    \State $\delta_\text{d} = \min\{\frac{\eta_\text{d} }{\Delta t}(E[k]-\underline{E})\,, 1\}$
    \vspace{0.25em}
    \If{$r[k] > 0$}
        \State $P_{\text{gen}}[k] \gets \min\{Cr[k], \overline{P}_{\!\text{gen}}\}$
        \State $P_{\text{CL}}[k] \gets 0$
        \State $P_{\text{batt}}^\text{c}[k] \gets 0$
        \State $P_{\text{batt}}^\text{d}[k] \gets \max\{0,\min\{Cr[k] - P_{\text{gen}}[k],\delta_\text{d}\overline{P}_{\text{batt}}\}\}$
        
    \Else
        \State $P_{\text{gen}}[k] \gets 0$
        \State $P_{\text{CL}}[k] \gets \min\{-Cr[k], \overline{P}_{\text{CL}}\}$
        \State $P_{\text{batt}}^\text{c}[k] \gets \min\{ 0, \max\{Cr[k] + P_{\text{CL}}[k], -\delta_\text{c}\overline{P}_{\text{batt}}\}\}$
        \State $P_{\text{batt}}^\text{d}[k] \gets 0$
        
    \EndIf
    \State $P_{\text{hes}}[k] \gets P_{\text{gen}}[k] - P_{\text{CL}}[k] + P^{\text{d}}_{\text{batt}}[k] + P^{\text{c}}_{\text{batt}}[k]$ 



\State \Return $P_{\text{hes}}[k]$ 
\end{algorithmic}
\end{algorithm}

\begin{theorem}
\label{thm:online_optimality}
    If, at optimality, the battery SoC constraints in~\eqref{eq:offline_opt} are not binding for every time step of the operational window, i.e.,
    \begin{equation}
        \underline{E} < E[k+1] < \overline{E}\,,~~\forall k \in \mathcal{K}\,,
    \end{equation}
    then the proposed real-time control strategy $g(C,r[k])$ (given by Algorithm~\ref{alg:hes_dispatch_short}) achieves optimal performance, i.e.,
    \begin{equation}
        J_\text{on}(C,\mathbf{r}) = J^{*}_\text{off}(C,\mathbf{r})\,,
    \end{equation}
    where
    \begin{equation}
        J_\text{on}(C,\mathbf{r}) \triangleq \sum_{k=0}^{N-1} \lvert Cr[k] - g(C,r[k]) \rvert\,.
    \end{equation}
\end{theorem}
\begin{IEEEproof}
    See Appendix. 
\end{IEEEproof}

Theorem~\ref{thm:online_optimality} provides conditions under which the tracking performance of the proposed real-time control strategy $J_\text{on}(C,\mathbf{r})$ is optimal (i.e., achieves the same as the performance of the offline dispatch), even though the power allocated to individual HES components may differ between the offline and real-time strategies (due to non-uniqueness of the offline dispatch). 
For a sufficiently large $C$, the battery SoC can saturate during the operational window, at which point, the proposed real-time control is not guaranteed to achieve optimal performance. 
However, based on historical hourly regulation signals, we find that the minimum performance threshold is typically violated before the battery hits SoC limits.

Thus, the proposed rule-based dispatch strategy allows straightforward real-time implementation, enabling the HES to track regulation signals near-optimally.
With this strategy defined, we can use historical regulation signal data to solve~\eqref{eq:cc_problem} and determine the optimal capacity bid as discussed in the next subsection.

\subsection{Determining the optimal HES capacity bid}
To determine the optimal capacity bid, we begin by assuming that we have $H$ samples of historical hours of the regulation signal, which we denote as $\mathbf{r}_i \, \forall \, i \in \{1,\dots,H\}$. 
For a given capacity bid $C$, we can then calculate the performance score, denoted $x^{g}_{\text{p},i}(C)$, that the proposed real-time control strategy $g(\cdot)$ would achieve for each regulation signal sample~$i$.
That is,
\begin{equation}
    x^{g}_{\text{p},i}(C) = 1 - \frac{\sum_{k=0}^{N-1} \lvert Cr_i[k] - g(C,r_i[k]) \rvert}
              {C \norm{\mathbf{r}_i}_1}\,,
\end{equation}
where $r_i[k]$ is the value of the $i$-th regulation signal sample at time step~$k$.

Given $C$, $g(\cdot)$, $\gamma$, and the performance score samples $x^{g}_{\text{p},i}(C)$, we define the probabilistic function $z^{g}_{\gamma}(C)$ which satisfies
\begin{equation}
    \mathbb{P}(x_\text{p} \ge z^{g}_{\gamma}(C)) = \gamma\,.
\end{equation}
For example, if $\gamma = 0.9$, then $z^{g}_{\gamma}(C)$ is the performance value such that exactly 90\% of the distribution of performance score samples (given $C$ and $g$) falls above $z^{g}_{\gamma}(C)$.
Next, we sweep through values of $C$, to find $\overline{C}$ such that $z^{g}_{\gamma}(\overline{C}) = \underline{x}_\text{p}$.
Consequentially, $\overline{C}$ is the maximum capacity bid that satisfies the chance-constraint~\eqref{eq:chance_constraint}.

\begin{figure}[tb]
\centering
\begin{subfigure}[b]{0.5\textwidth}
\centering
\includegraphics[width=\textwidth]{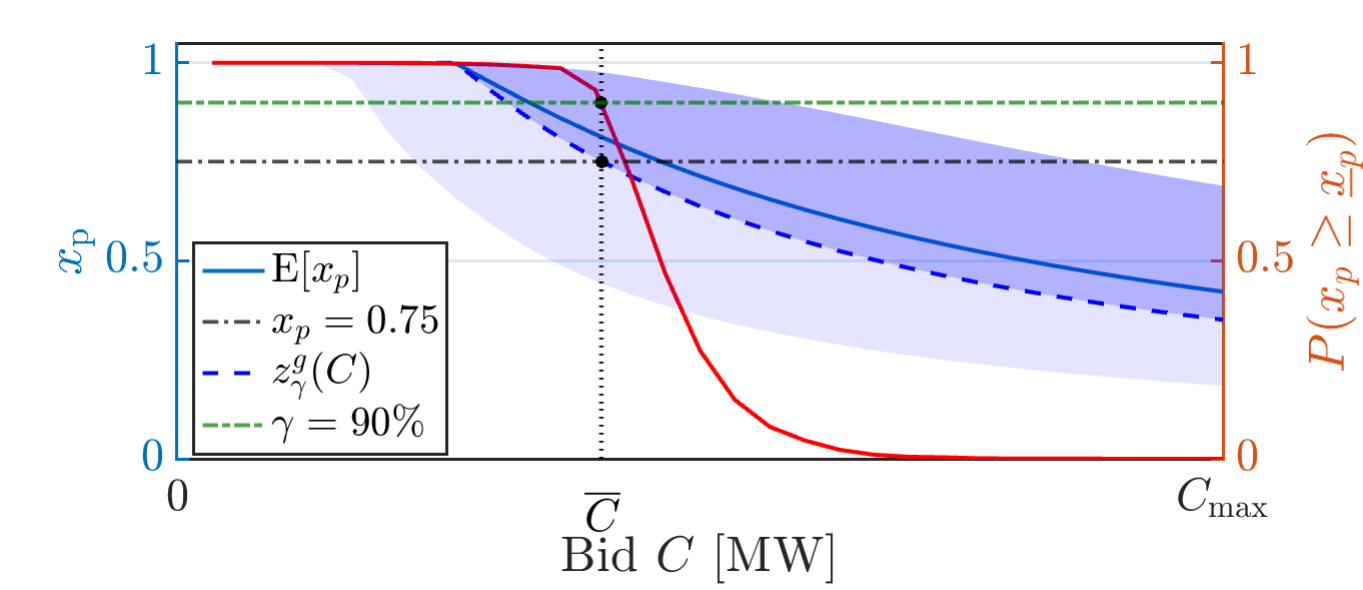}
\caption{$C$ vs $x_{\text{p}}$}
\label{fig:P_gen_variation}
\end{subfigure}
\begin{subfigure}[b]{0.5\textwidth}
\centering
\includegraphics[width=\textwidth]{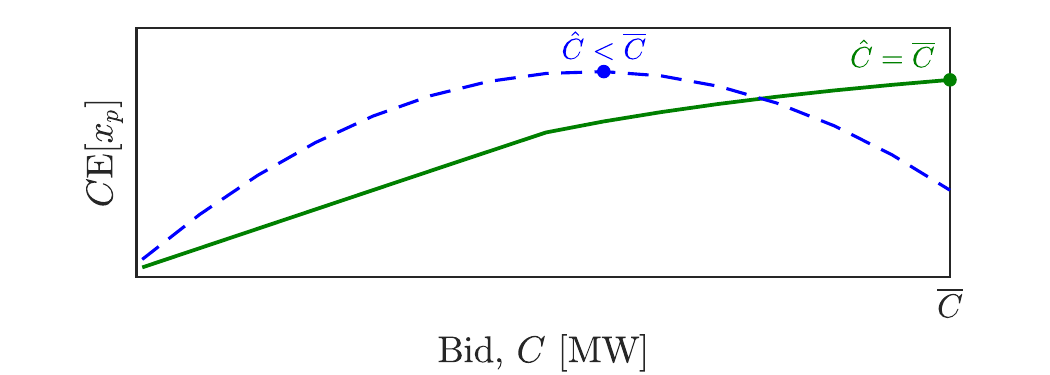}
\caption{$C$ vs $C\mathbb{E}[x_{\text{p}}]$}
\label{fig:P_CL_variation}
\end{subfigure}
\caption{Illustration of distribution of $x_{\text{p}}$ and $C\mathbb{E}[x_{\text{p}}]$ across various bids $C$ for a given control strategy $g(\cdot)$. The optimal bid $C^{*} = \min\{\hat{C}, C_{\text{max}}\}$ is selected to satisfy the chance constraint $\mathbb{P}(x_{\text{p}} \ge \underline{x}_{\text{p}}) \ge \gamma$ i.e., $z_{\gamma}^{g}(\overline{C}) = \underline{x}_{\text{p}}$, while maximizing expected revenue. 
In (a), the shaded regions show the support of the distribution of $x_{\text{p}}$, where the darker region shows that $\gamma=90\%$ of the distribution falls above $z_{\gamma}^{g}(C)$. }
\label{fig:det_cap_bid}
\vspace{-0.3cm}
\end{figure}

Finally, we can use the performance score samples $x^{g}_{\text{p},i}(C)$ to compute empirical approximations of $\mathbb{E}[x_\text{p}]$ for different values of $C$.
More specifically, we find $\hat{C}\in(0,\overline{C}]$ which results in the maximum value of $C\mathbb{E}[x_\text{p}]$.
The optimal capacity bid $C^*$ is then given by:
\begin{equation}
    C^* = \min\{\hat{C}\,,C_\text{max}\}\,.
\end{equation}
This process for determining the optimal capacity bid is shown in Fig.~\ref{fig:det_cap_bid}.
In the next section, we evaluate the performance of the proposed capacity bidding and real-time HES control strategies using a case study.

\section{Case Study: Frequency Regulation Market}\label{case_studies}
Among various ISOs and utilities, PJM operates one of the most efficient regulation markets, with a high regulation-to-generation ratio and a near energy-neutral Reg-D signal, which makes it particularly suitable for battery-based HES \cite{Bolun2016}. Therefore, we select the PJM Reg-D signal to test the proposed bidding and dispatch approach~\cite{pjm2025manual12}. We first characterize the historical PJM Reg-D signal, a fast-response regulation signal, then test the proposed optimal bidding and real-time control strategies for both symmetrical ($\overline{P}_{\text{gen}}$ = $\overline{P}_{\text{CL}}$) and asymmetrical ($\overline{P}_{\text{gen}} \neq \overline{P}_{\text{CL}}$) HES configurations, as discussed in the subsequent subsections. 

\subsection{Statistical characterization of Reg-D signal}
To characterize the Reg-D signal, we used one year of historical data sampled at 2-second intervals. The signal is analyzed in terms of its energy neutrality, $W$,  and worst-case drift, ${W}_{\infty}$, as defined by 
\begin{align*}
    W &:= \sum_{k=0}^{N-1} r[k] \, \Delta t  \qquad 
    W_{\infty} &:= \max_{j\in[0,N-1]} \left| \sum_{k=0}^{j} r[k] \, \Delta t \right| . \label{eq_Ws}
\end{align*}

\begin{figure}[tb]
\centering
\begin{subfigure}[b]{0.4\textwidth}
\centering
\includegraphics[width=\textwidth]{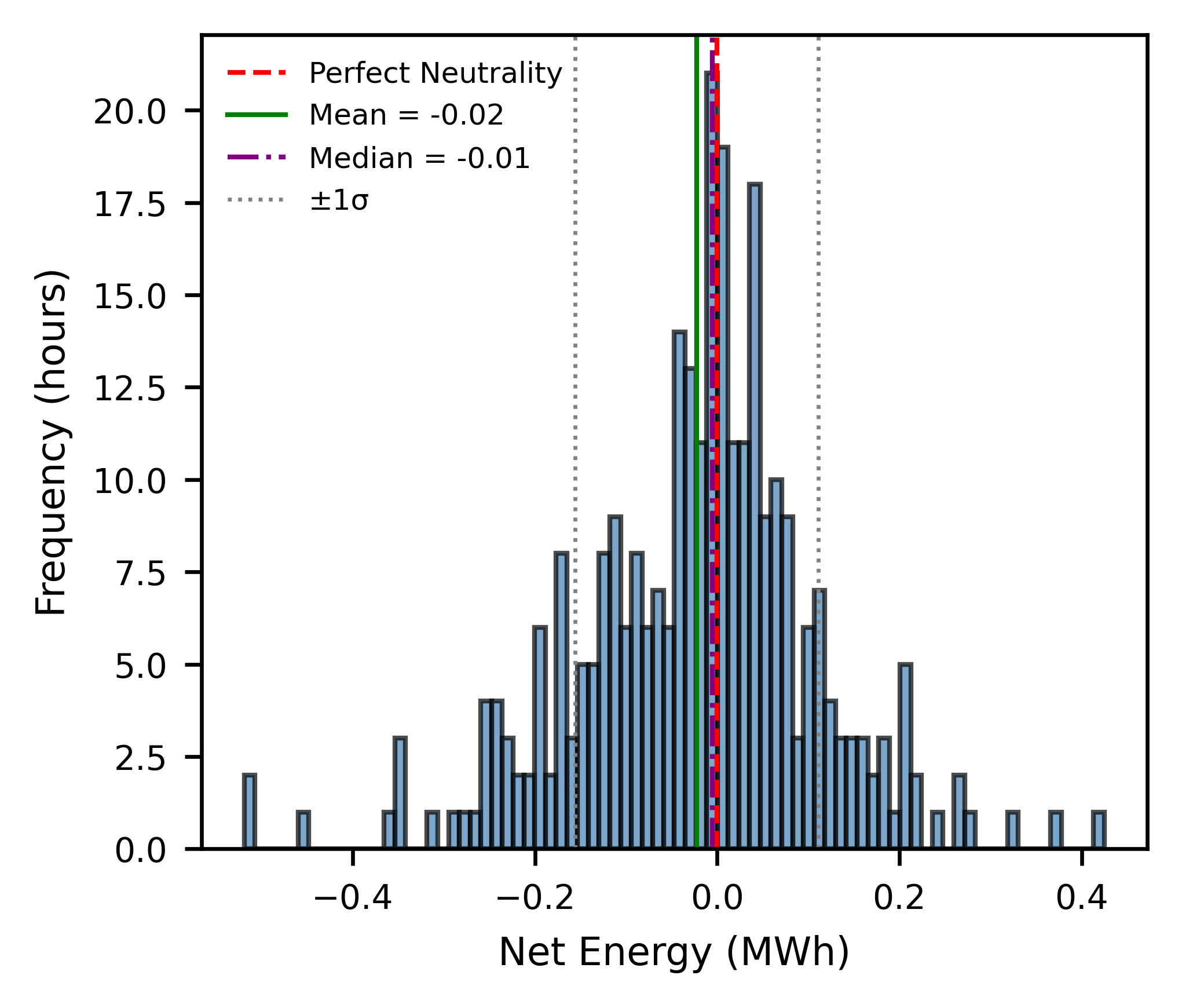}
\caption{Distribution of hourly energy neutrality, $W$.}
\label{}
\end{subfigure}
\begin{subfigure}[b]{0.4\textwidth}
\centering
\includegraphics[width=\textwidth]{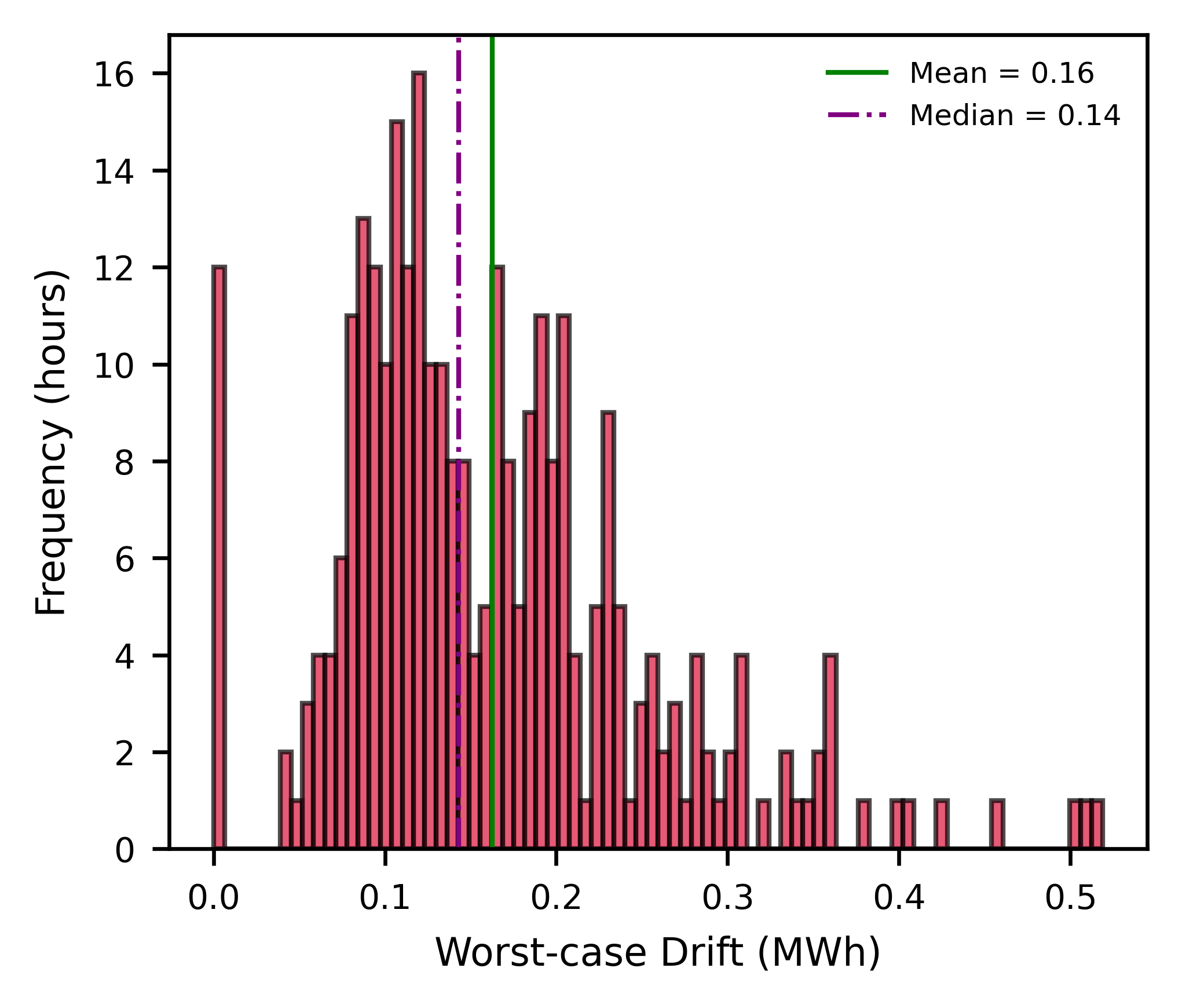}
\caption{Distribution of worst case energy drift, $W_\infty$.}
\label{}
\end{subfigure}
\caption{Statistical characterization of the PJM Reg-D signal over one year: (a) distribution of hourly energy drift, showing deviations from perfect neutrality; (b) distribution of worst case drift for normalized regulation signal.}
\label{Rev_comp2}
\end{figure}

Here, $W$ characterizes the net energy content of the Reg-D signal over a one-hour horizon, with mean $\tilde{r} = -0.02$ MWh, confirming that the signal is nearly energy-neutral.
 The maximum absolute cumulative deviation of the regulation signal is quantified by $W_{\infty}$. 
 In the historical data, $W_{\infty}$ reaches approximately half of $C^{*}$ MWh for the optimal bid $C^{*}$, but only during a few extreme hours. 
This implies that a battery-only resource would experience significant SoC excursions if dispatched independently. This observation underscores the need for coordinated use of the generator and controllable load, which can alleviate SoC drift and enable reliable long term regulation performance. Fig.~\ref{Rev_comp2} shows the distributions of these metrics over one year. 
This statistical characterization provides insight into the operational requirements and stresses imposed on the HES, and serves as a foundation for estimating $\mathbb{E}[x_{\text{p}}]$ and revenue $R$ under different bidding strategies.

\begin{table}[!t]
\centering
\caption{HES Parameters}
\begin{tabular}{l c}
\hline
\textbf{Component} & \textbf{Value} \\
\hline
Generator rated power $\overline{P}_{\!\text{gen}}$ & 3 MW \\
Load rated power $\overline{P}_{\text{CL}}$ & 3 MW \\
Battery rated power $\overline{P}_{\text{batt}}$& 5 MW \\
Battery energy capacity & 5 MWh \\
Battery round-trip efficiency $\eta_\text{d} =\eta_\text{c} $ & 95\% \\
Initial SoC  $E_{0}$ & 0.50 p.u.\\
Max SoC  $\overline{E}$ & 0.90 p.u.\\
Min SoC  $\underline{E}$ & 0.10 p.u.\\
\hline
\end{tabular}
\label{tab:HES_params}
\end{table}
\subsection{Performance evaluation for symmetrical HES}
Table \ref{tab:HES_params} summarizes the HES parameters, providing a symmetric capacity of 8 MW for both up and down regulation. The battery state of charge is constrained between 0.1 and 0.9, and we assume the initial battery SoC $E_0 = 0.5$ to start from a neutral point and avoid bias toward charging or discharging. These settings ensure that the battery, generator, and controllable load can collectively track the regulation signal while respecting their operational limits.

\begin{figure}[t]
\centering
\includegraphics[width=1.0\linewidth]{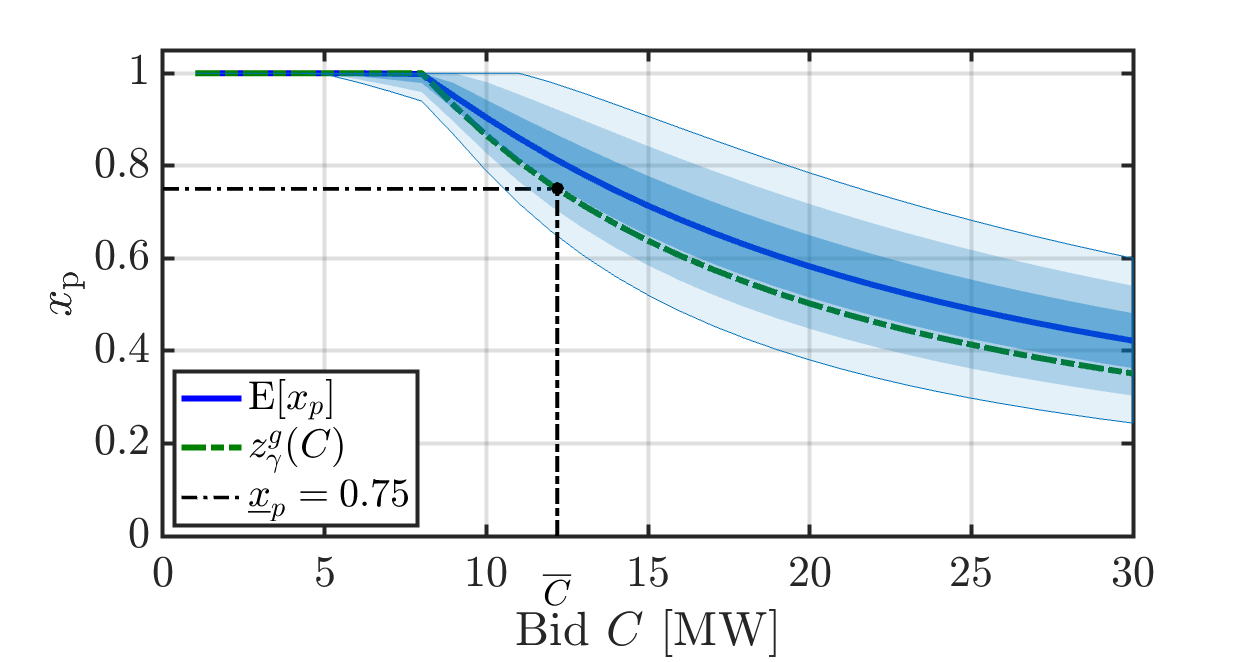}
\caption{Distribution of $x_{\text{p}}$ across different bid capacities under the real-time control strategy. The shaded areas show $\pm\sigma$, $\pm2\sigma$, and $\pm3\sigma$ standard deviation bands.}
\label{fig:HES_Performance}
\vspace{-0.3cm}
\end{figure}

Further, Table~\ref{tab:hes_performance} summarizes the HES performance for selected bid capacities. For each bid, we calculate the mean performance index $\mathbb{E}[x_\text{p}]$, as well as the probability of meeting the minimum performance threshold $\mathbb{P}(x_\text{p} \ge \underline{x}_\text{p})$, where $\underline{x}_\text{p} = 0.75$. As expected, smaller bids consistently achieve near-perfect performance, while higher bids lead to a gradual reduction in both $\mathbb{E}[x_\text{p}]$ and the probability of compliance due to the limited capacity of the HES assets. This table complements the $x_{\text{p}}$ distribution plot shown in Fig.~\ref{fig:HES_Performance} by providing exact numerical values for key bid points, showing how the HES can reliably participate in the market under varying bid sizes.
\begin{table}[tb]
\centering
\caption{HES Performance Summary for Selected Capacity Bids}
\label{tab:hes_performance}
\begin{tabular}{c c c c}
\hline
$C$ [MW] & $\mathbb{E}[x_\text{p}]$ & $z^{g}_{\gamma}(C)$ &$\mathbb{P}(x_\text{p} \ge 0.75)$ \\
\hline
1  & 1.000  & 1.000 & 1.000 \\
5  & 1.000  & 1.000 & 1.000 \\
8  & 0.999  &0.999 & 1.000\\ 
10 & 0.905  &0.865 & 0.997 \\
\rowcolor{green!10}
\fbox{12.21} & 0.813  & \fbox{0.750} & \fbox{0.900} \\
15 & 0.715  &0.637  & 0.273 \\
20 & 0.583 & 0.502 & 0.010 \\
\hline
\end{tabular}
\end{table}

Using PJM Reg-D signals as training data and the proposed real-time HES control strategy, the maximum capacity bid satisfying a 90\% confidence threshold (i.e., $\gamma = 0.9$) was found to be $C^{*} = 12.21$~MW using \eqref{Bid_optimization}. 
The expected performance score at this bid is $\mathbb{E}[x_{\text{p}}] = 0.812$, and the expected revenue $R$ is 9.9$(\lambda_{\text{C}} +\lambda_{\text{M}} M)$, which is 24\% higher than the profit that would be obtained by bidding the HES capacity (8~MW) with perfect performance ($x_{\text{p}} = 1.0$, $\gamma = 1$). These results demonstrate that higher bids increase potential revenue, but reduce the confidence in achieving the minimum required performance.

\begin{figure}[!htb]
\centering
\begin{subfigure}[b]{0.48\textwidth}
\centering
\includegraphics[width=\textwidth]{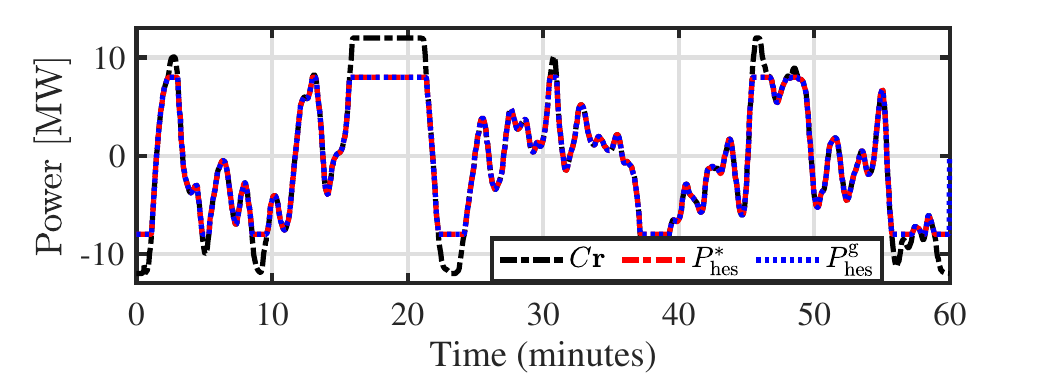}
\caption{Regulation signal and HES power}
\label{}
\end{subfigure}
\begin{subfigure}[b]{0.48\textwidth}
\centering
\includegraphics[width=\textwidth]{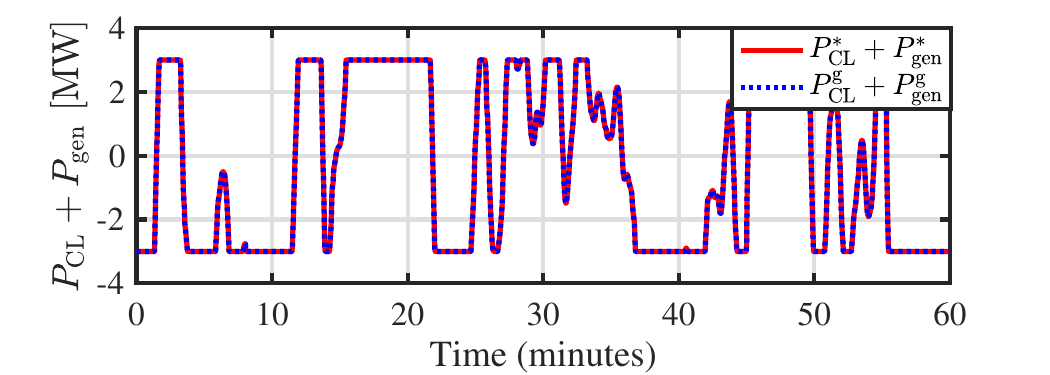}
\caption{Net controllable generator and load: $P_{\text{gen}}$+$P_{\text{CL}}$}
\label{}
\end{subfigure}
\begin{subfigure}[b]{0.48\textwidth}
\centering
\includegraphics[width=\textwidth]{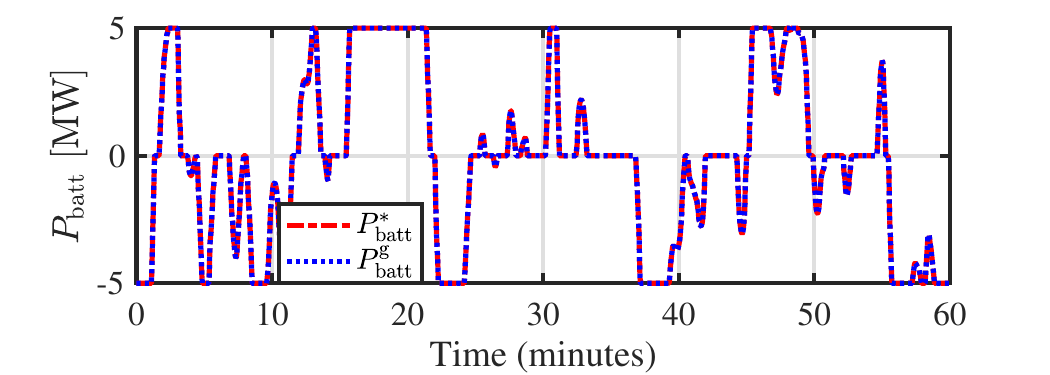}
\caption{Battery net power $P_{\text{batt}}$}
\label{}
\end{subfigure}
\begin{subfigure}[b]{0.48\textwidth}
\centering
\includegraphics[width=\textwidth]{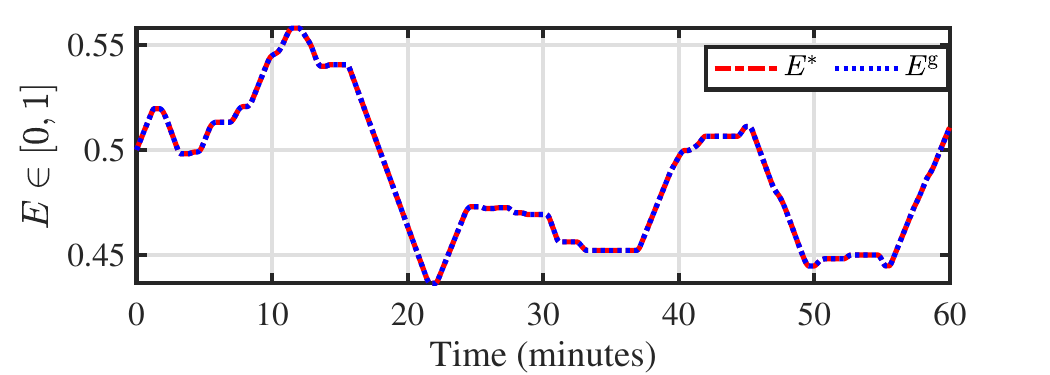}
\caption{Battery SoC $E$}
\end{subfigure}
\caption{Comparison of real-time dispatch $g(\cdot)$ and optimal offline dispatch (denoted with a $^*$) for one representative hour with $C=12.21$ MW. The plots show the net regulation signal $C\mathbf{r}$, $P_{\text{hes}}$, $P_{\text{gen}}$+ $P_{\text{CL}}$, $P_{\text{batt}}$,  and state of charge $E$. The performance $x_{\text{p}}=0.8412$ for both strategies.}
\label{fig:Power Allocation}
\end{figure}

Using the optimal bid $C^{*} = 12.21$~MW, we implemented the real-time dispatch strategy $g(\cdot)$ from Algorithm~\ref{alg:hes_dispatch_short}, and benchmarked it against the offline optimal dispatch~\eqref{eq:offline_opt}. 
Fig.~\ref{fig:Power Allocation}(a) compares the HES power outputs from the offline optimization and the real-time control strategy as they track the scaled regulation signal $C\mathbf{r}$. 
Figs.~\ref{fig:Power Allocation}(b) and~\ref{fig:Power Allocation}(c) compare the allocation of power among HES components between the two strategies, and Fig.~\ref{fig:Power Allocation}(d) shows the evolution of battery SoC. 
While there are non-unique optimal allocations between generation and controllable load, the combined contribution $P_{\text{gen}} + P_{\text{CL}}$ remains the same for both policies. 
The performance indices achieved for both the offline optimal dispatch and the real-time control strategy are identical, where $x_{\text{p}} = 0.841$.
This results in $28\%$ higher revenue compared to just bidding the maximum HES capacity (i.e. 8~MW) at $x_{\text{p}} \approx 1$.  

This comparison demonstrates that the proposed real-time strategy $g(\cdot)$ achieves the same performance as the offline optimal dispatch strategy~\eqref{eq:offline_opt} while remaining implementable in real time. 
It effectively balances battery, generator, and load contributions to follow the regulation signal. The results also highlight the economic trade off in capacity bidding i.e. bidding higher than the capacity of HES can yield higher revenue when performance is maintained, but excessive overbidding leads to lower performance and potential penalties. Therefore, the real-time control strategy provides a practical, high-performing approach for HES participation in the regulation market.

\subsection{Performance evaluation for asymmetrical HES}
In the previous subsection, a symmetric HES with $\overline{P}_{\text{gen}}~=~\overline{P}_{\text{CL}}$ was analyzed, providing equal export and import capability for regulation tracking. Unlike standalone storage, the HES is not constrained by symmetric power limits. Hence, this section discusses asymmetric cases by varying $\overline{P}_{\text{gen}}$ and $\overline{P}_{\text{CL}}$ to assess their impact on bid capacity and regulation performance. 

Three asymmetric HES configurations are analyzed and compared against the symmetric case:
\begin{itemize}
    \item \textbf{Case~I}: In the first case, the HES operates with either $\overline{P}_{\text{CL}}=0$ (no load) or $\overline{P}_{\text{gen}}=0$ (no generation). 
    \item 
    \textbf{Case~II}: The second case sets the maximum controllable load or generator capacity equal to the battery limit (i.e., $\overline{P}_{\text{CL}}=\overline{P}_{\text{batt}}$ or $\overline{P}_{\text{gen}}=\overline{P}_{\text{batt}}$), while the remaining asset is set to its original limit as shown in Table~\ref{tab:HES_params} (i.e., $\overline{P}_{\text{gen}}$ or $\overline{P}_{\text{CL}}=3~\mathrm{MW}$).
    \item \textbf{Case~III}: The third case sets $\overline{P}_{\text{CL}}=\overline{P}_{\text{batt}}+\overline{P}_{\text{gen}}$ (more load) or $\overline{P}_{\text{gen}}=\overline{P}_{\text{batt}}+\overline{P}_{\text{CL}}$ (more generation).
\end{itemize}

\begin{figure}[!htb]
\centering
\begin{subfigure}[b]{0.5\textwidth}
\centering
\includegraphics[width=\textwidth]{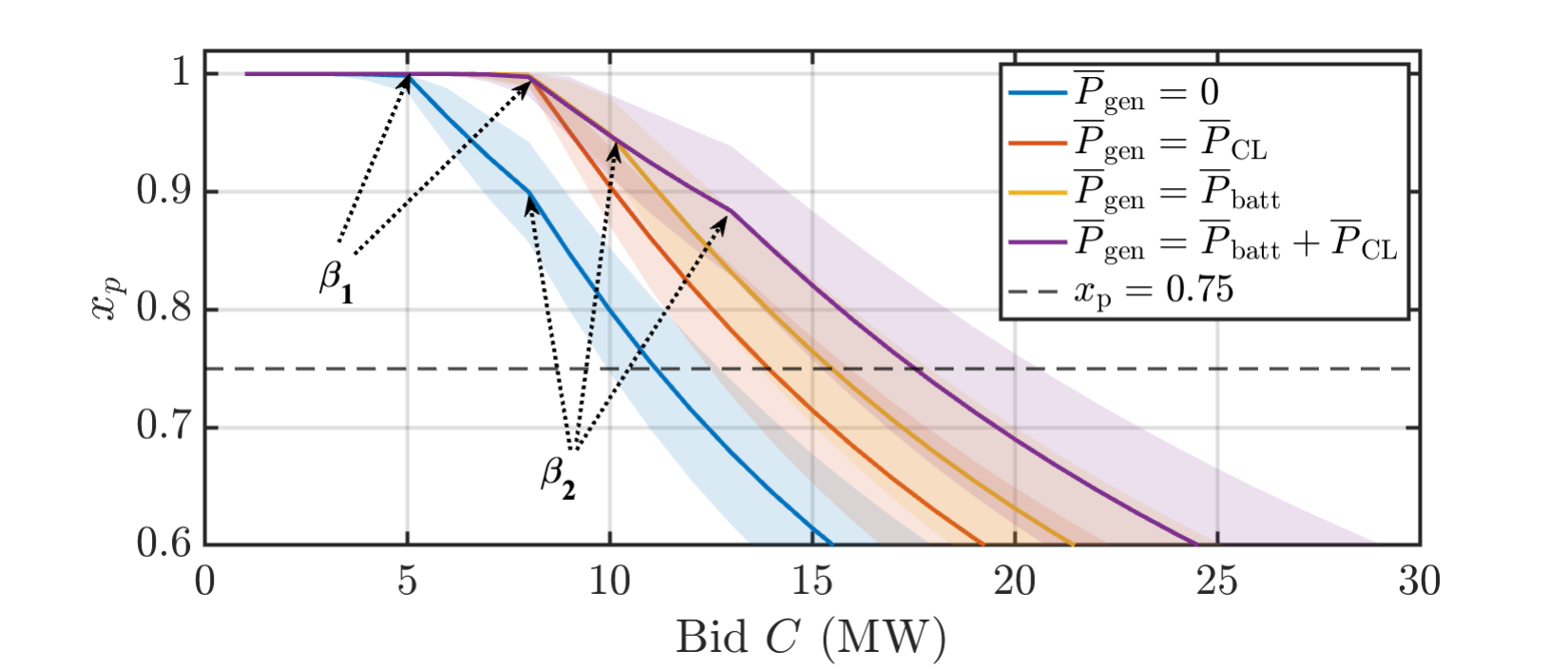}
\caption{Variation in generator capacity ($\overline{P}_{\text{gen}}$) while $\overline{P}_{\text{CL}}$ and $\overline{P}_{\text{batt}}$ are fixed.}
\label{fig:P_gen_variation}
\end{subfigure}
\begin{subfigure}[b]{0.5\textwidth}
\centering
\includegraphics[width=\textwidth]{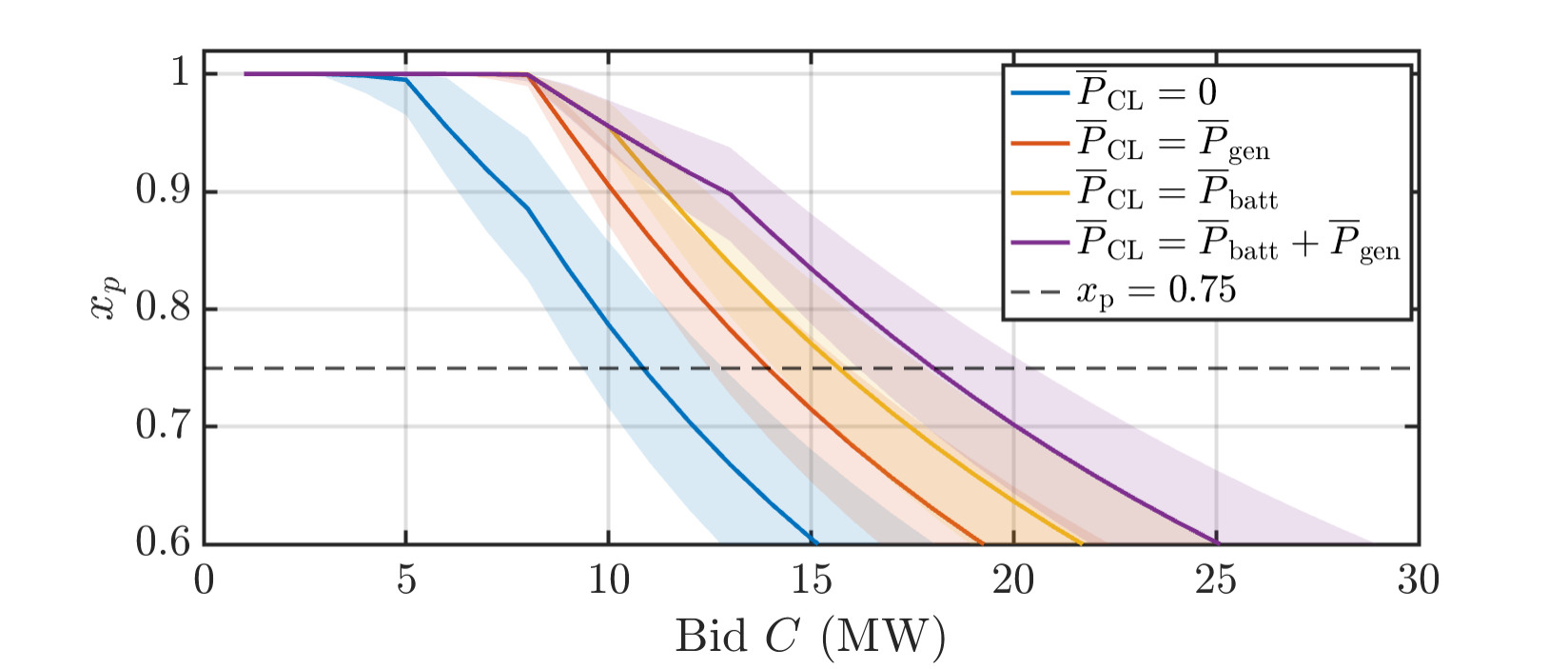}
\caption{Variation in controllable load capacity ($\overline{P}_{\text{CL}}$) while $\overline{P}_{\text{gen}}$ and $\overline{P}_{\text{batt}}$ are fixed.}
\label{fig:P_CL_variation}
\end{subfigure}
\caption{Comparison of the mean performance index $x_\text{p}$ as a function of capacity bid $C$ for varying controllable load and generator power capacities. The shaded bands represent one standard deviation ($\pm \sigma$).}
\label{fig:Asymmetric_HES}
\vspace{-0.3cm}
\end{figure}

The performance curve of the asymmetric HES is monotonic with two knee points, as shown in Fig.~\ref{fig:Asymmetric_HES}. It remains nearly constant up to the first knee point, $\beta_1 = \min \{(\overline{P}_{\text{gen}} + \overline{P}_{\text{batt}}), (\overline{P}_{\text{CL}} + \overline{P}_{\text{batt}})\}$, which represents the \textit{best-performance region} ($x_{\text{p}} \approx 1$). Beyond this point, the performance gradually declines within the \textit{partial-performance region} until the second knee point, $\beta_2 = \max \{(\overline{P}_{\text{gen}} + \overline{P}_{\text{batt}}), (\overline{P}_{\text{CL}} + \overline{P}_{\text{batt}})\}$, corresponding to the maximum attainable capacity on either side. Past $\beta_2$, the system performance deteriorates sharply. In the case of a symmetric HES, $\beta_1 = \beta_2$, resulting in only two distinct performance regions. 
Table~\ref{tab:asym_hes_performance} shows the correlation between increase in power capacity of the assets and the optimal capacity bid. The optimal bid initially increases with asset capacity but eventually saturates as the generator or load becomes excessively large. This trend occurs because the battery reaches its energy limits when it can participate only in one direction (charging or discharging).


\begin{table}[tb]
\centering
\caption{Optimal Capacity Bids for Asymmetric HES}
\label{tab:asym_hes_performance}
\begin{tabular}{c c c|c c c}
\hline
$\overline{P}_{\text{gen}}$ [MW] & $C^{*}$ & $\mathbb{E}[x_\text{p}]$ & $\overline{P}_{\text{CL}}$ [MW] &$C^{*}$ & $\mathbb{E}[x_\text{p}]$\\
\hline
0  & 9.78 &0.810 & 0 & 9.2& 0.824 \\
3  & 12.21 &0.812 & 3 & 12.21 &0.812\\
5 & 13.43 & 0.816 & 5 & 13.75& 0.810 \\ 
8 & 14.88 & 0.824 &8 & 15.8 & 0.812\\
13 & 15.78  &0.833 &13 &18.12  &0.821\\
25 & 14.92  &0.844 &25 &17.34  &0.828\\
50 & 14.92  & 0.844&50 & 17.34 & 0.828\\
\hline
\end{tabular}
\end{table}

\begin{figure}[t]
\centering
\begin{subfigure}[b]{0.5\textwidth}
\centering
\includegraphics[width=\textwidth]{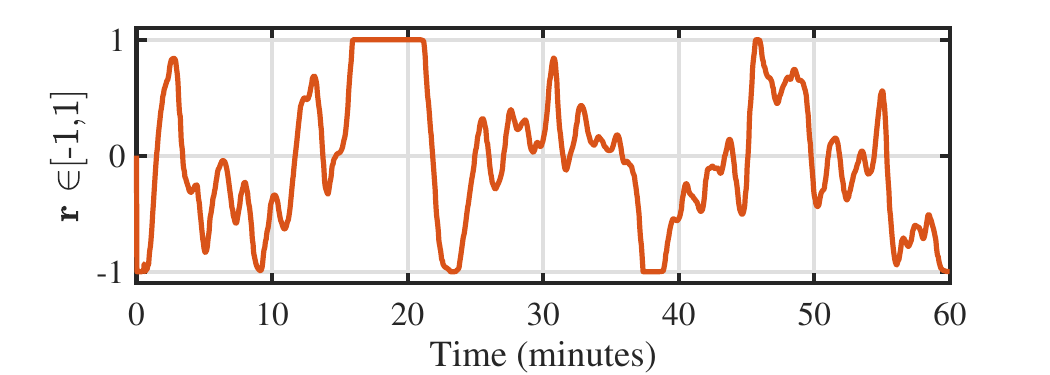}
\caption{Reg-D signal.}
\label{fig:P_gen_variation}
\end{subfigure}
\begin{subfigure}[b]{0.5\textwidth}
\centering
\includegraphics[width=\textwidth]{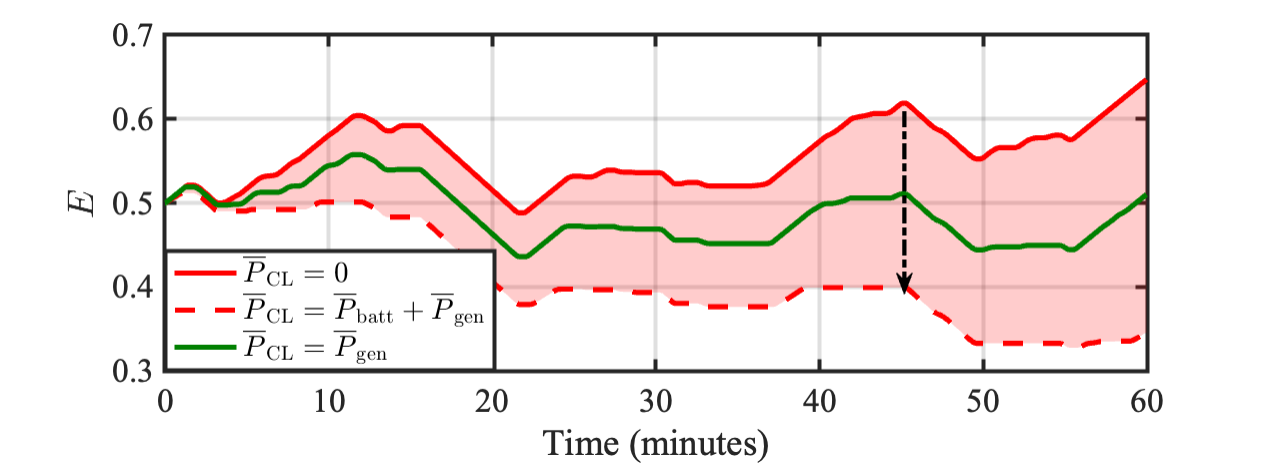}
\caption{Variation in controllable load capacity ($\overline{P}_{\text{CL}}$) while $\overline{P}_{\text{gen}}$ and $\overline{P}_{\text{batt}}$ are fixed.}
\label{fig:P_gen_SoC_variation}
\end{subfigure}
\begin{subfigure}[b]{0.5\textwidth}
\centering
\includegraphics[width=\textwidth]{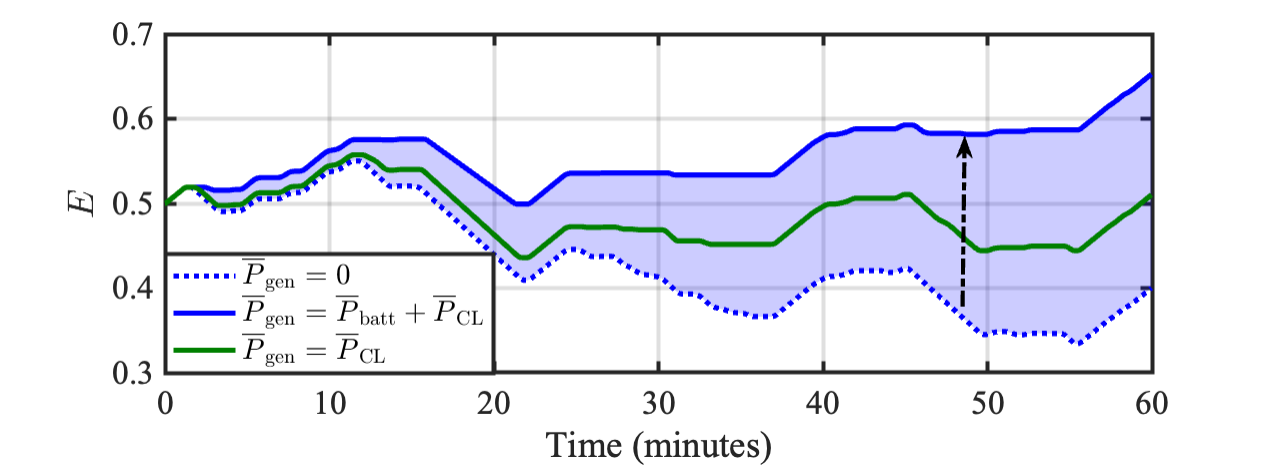}
\caption{Variation in generator capacity ($\overline{P}_{\text{gen}}$) while $\overline{P}_{\text{CL}}$ and $\overline{P}_{\text{batt}}$ are fixed.}
\label{fig:P_CL_SoC_variation}
\end{subfigure}
\caption{SoC drift with respect to variations in $\overline{P}_{\text{CL}}$ and $\overline{P}_{\text{gen}}$, while $\overline{P}_{\text{batt}}$ remains fixed. The arrows indicate the direction of SoC drift as the corresponding power limits increase from zero for a given Reg-D signal $\mathbf{r}$.}
\label{fig:Asymmetric_SoC}
\end{figure}

\begin{figure}[t]
\centering
\includegraphics[width=1.0\linewidth]{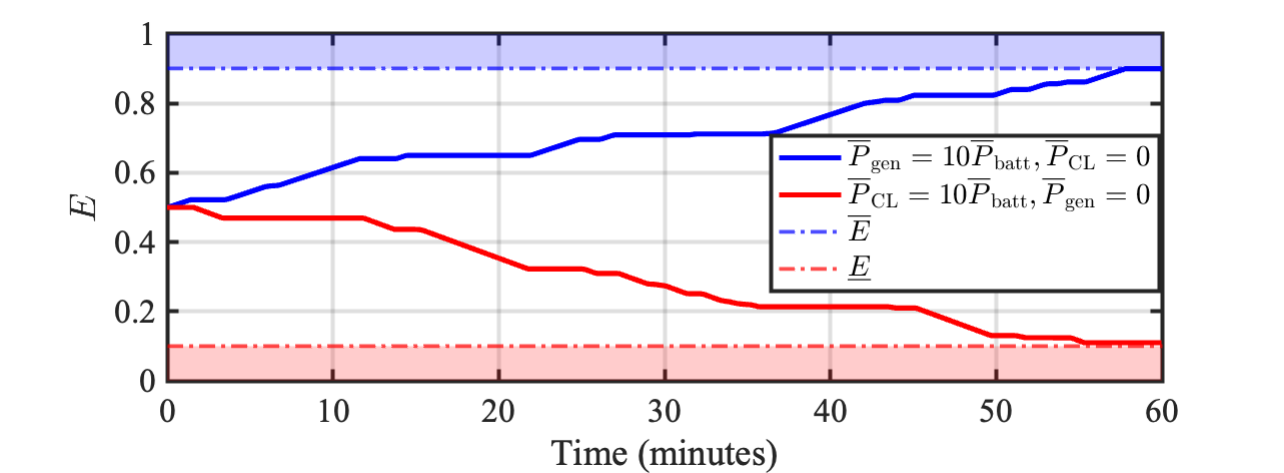}
\caption{SoC response under the Reg-D signal $\mathbf{r}$ when either ${\overline{P}}_{\text{CL}}$ or ${\overline{P}}_{\text{gen}}$ is set to ten times $\overline{P}_{\text{batt}}$, while the other asset is inactive. In both cases, the SoC hit the limits.}
\label{fig:Asymmetric_SoC_2}
\vspace{-0.3cm}
\end{figure}

Since the performance begins to degrade sharply once the bid exceeds the minimum power limit in either direction, this raises an important question: how large can these two assets be scaled without imposing excessive strain on the battery? 
Fig.~\ref{fig:Asymmetric_SoC} presents the SoC drift with respect to variations in $\overline{P}_{\text{CL}}$ and $\overline{P}_{\text{gen}}$, while $\overline{P}_{\text{batt}}$ remains fixed under the given Reg-D signal~$\mathbf{r}$. This ensures that one side of the system retains the same total capacity while the other asset is varied. In this case, the bid capacity was kept the same as that of the symmetric HES, i.e., $C~=~12.21~\mathrm{MW}$, to have a fair comparison between the symmetric and asymmetric HES. Further, Fig.~\ref{fig:Asymmetric_SoC_2} shows the SoC response when either $\overline{P}_{\text{CL}}$ or $\overline{P}_{\text{gen}}$ is set to ten times $\overline{P}_{\text{batt}}$ while the other asset is inactive. The bid was varied to identify the point at which the SoC limits are reached, and in both cases, the SoC eventually hits its operational bounds for $C~=~20~\mathrm{MW}$, confirming that excessive asymmetry can lead to sustained energy imbalance and battery saturation.


These observations highlight the critical role of the HES configuration: while the non-battery assets enhance dispatch flexibility and bidding potential, maintaining balanced power limits across all HES components is essential to ensure reliable performance.

\section{Conclusion}
\label{Conclusion}
This paper presented a unified framework for performance-based participation of HES in frequency regulation markets. A chance-constrained bidding formulation was proposed to determine the maximum feasible bid $C^{*}$ that ensures the performance index $x_{\text{p}} \ge \underline{x}_{\text{p}}$ with high statistical confidence $\gamma$. Further, the proposed real-time control strategy $g(\cdot)$ allocates power among the generator, battery, and controllable load in a way that respects operational constraints while achieving identical performance to offline optimal dispatch (when SoC limits are not binding).

Simulation results using PJM Reg-D data demonstrate the trade-off between performance and profitability. Although higher bids increase expected revenue (proportional to $C\mathbb{E}[x_{\text{p}}]$), they also increase the risk of performance violations. The optimal bid $C^{*} = \min\{\hat{C}, C_{\text{max}}\}$ satisfies $\mathbb{P}(x_{\text{p}} \ge \underline{x}_{\text{p}}) \ge \gamma$, ensuring both profitability and compliance. Furthermore, analysis of symmetric and asymmetric HES configurations showed that asymmetry introduces a performance curve with two knee points ($\beta_1$, $\beta_2$) defining distinct operational regions. The asymmetry also creates a drift in battery SoC toward a lower power limit component, offering insight into the limits of flexible capacity.

The proposed framework provides a practical and computationally efficient foundation for integrating stochastic bidding and real-time control in market participation. Future extensions will focus on incorporating intermittent resources (e.g., PV, wind) and flexible loads (e.g., EV chargers, data centers), enabling coordinated participation of multiple HES units to enhance grid flexibility and regulation reliability.

\appendix
\label{appendix}
\vspace{-0.2cm}
\subsection*{Proof of Theorem~\ref{thm:online_optimality}}

We prove Theorem~\ref{thm:online_optimality} via Karush-Kuhn-Tucker (KKT) analysis.
First, we consider the offline optimal dispatch problem~\eqref{eq:offline_opt}.
This problem is non-convex due to the complementarity constraint~\eqref{eq:batt_cc}; thus, the KKT analysis provides only necessary conditions for global optimality, not sufficient.
To address this, we consider a convex relaxation of~\eqref{eq:offline_opt} by removing the constraint~\eqref{eq:batt_cc}.
This convex problem provides a lower bound on the optimal objective function value, i.e.,
\begin{equation}
    J^{*}_\text{cvx}(C,\mathbf{r}) \le J^{*}_\text{off}(C,\mathbf{r})\,,
\end{equation}
where $J^{*}_\text{cvx}(C,\mathbf{r})$ denotes the value of the objective function for the convex problem at optimality.

Additionally, since the absolute value in the objective function is not differentiable, we reformulate the convex relaxation of~\eqref{eq:offline_opt} into the following equivalent problem by introducing an auxiliary variable~$w[k]$:
\begin{subequations}
\label{eq:convex_prob}
\label{eq:offline_opt_reform}
\begin{align}
\min_{\mathbf{y}}~~&\sum_{k=0}^{N-1} w[k]\,,  \\
\text{s.t.}~~&-w[k] \le Cr[k] - P_{\text{hes}}[k] \le w[k]\,,~~\forall k \in \mathcal{K}\,,\label{eq:w_lims}\\
&\eqref{hes_power}\text{--}\eqref{eq:batt_c_lim}~\text{and}~ \eqref{eq:batt_dyn}\text{--}\eqref{eq:batt_soc_lim}\,.
\end{align}
\end{subequations}

Next, we form the Lagrangian $\mathcal{L}(\mathbf{y},\boldsymbol{\lambda},\boldsymbol{\nu})$ and derive the following stationarity conditions, which hold for all $k\in\mathcal{K}$:
\begin{subequations}
\begin{align}
    \nabla_{w[k]}\mathcal{L} = 0 \implies& \underline{\lambda}_{w}[k] + \overline{\lambda}_{w}[k] = 1\,,\label{eq:stat_lam_w}\\
    \nabla_{P_\text{hes}[k]}\mathcal{L} = 0 \implies& \underline{\lambda}_{w}[k] - \overline{\lambda}_{w}[k] + \nu_\text{hes}[k] = 0\,,\label{eq:nu_hes}\\
    \nabla_{P_\text{gen}[k]}\mathcal{L} = 0 \implies& - \underline{\lambda}_\text{gen}[k] +  \overline{\lambda}_\text{gen}[k] = \nu_\text{hes}[k]\,,\label{eq:stat_gen}\\
    \nabla_{P_\text{CL}[k]}\mathcal{L} = 0 \implies& - \underline{\lambda}_\text{CL}[k] +  \overline{\lambda}_\text{CL}[k] = -\nu_\text{hes}[k]\,,\label{eq:stat_load}\\
    \nabla_{P_\text{batt}^{\text{d}}[k]}\mathcal{L} = 0 \implies& \nonumber\\
     &\hspace{-2.5cm}-\underline{\lambda}_\text{batt}^\text{d}[k] +  \overline{\lambda}_\text{batt}^\text{d}[k]+\frac{\Delta t}{\eta_\text{d}\overline{P}_\text{batt}}\nu_{E}[k] = \nu_\text{hes}[k]\,,\label{eq:stat_discharge}\\
    \nabla_{P_\text{batt}^{\text{c}}[k]}\mathcal{L} = 0 \implies& \nonumber\\
     &\hspace{-2.5cm}-\underline{\lambda}_\text{batt}^\text{c}[k] +  \overline{\lambda}_\text{batt}^\text{c}[k]+\frac{\eta_\text{c}\Delta t}{\overline{P}_\text{batt}}\nu_{E}[k] = \nu_\text{hes}[k]\,,\label{eq:stat_charge}\\
     \nabla_{E[k+1]}\mathcal{L} = 0 \implies \nonumber \\
     &\hspace{-3.0cm}
     \nu_{E}[k] = 
     \begin{cases}
        \nu_{E}[k+1]- \underline{\lambda}_{E}[k] +\overline{\lambda}_{E}[k]\,,\forall k \in \mathcal{K}\setminus\{N-1\}\,, \\
        - \underline{\lambda}_{E}[k] +\overline{\lambda}_{E}[k]\,,\hspace{1.5cm}k = N-1\,.
      \end{cases}\label{eq:KKT_recursion}
\end{align}
\end{subequations}
Here, $\underline{\lambda}_{w}[k]$ and $\overline{\lambda}_{w}[k]$ are non-negative dual variables associated with the lower and upper bounds in~\eqref{eq:w_lims}.
Similarly, other $\lambda$ dual variables are associated with the inequality constraints on controllable generation, load, battery charging and discharging in~\eqref{eq:gen_pow_lim}--\eqref{eq:batt_c_lim} and battery SoC in~\eqref{eq:batt_soc_lim}.
The dual variables $\nu_\text{hes}[k]$ and $\nu_{E}[k]$ correspond to the equality constraints~\eqref{hes_power} and~\eqref{eq:batt_dyn}, respectively.

Under the assumption of Theorem~\ref{thm:online_optimality} (i.e., battery SoC bounds are not binding for all $k\in \mathcal{K}$),
the complementary slackness constraints associated with $\underline{\lambda}_{E}[k]$ and $\overline{\lambda}_{E}[k]$ will force these dual variables to zero.
Furthermore, via recursion on~\eqref{eq:KKT_recursion}, we have $\nu_{E}[k]=0$ for all $k\in\mathcal{K}$.
Thus,~\eqref{eq:stat_discharge} and~\eqref{eq:stat_charge} simplify to
\begin{align}
    &-\underline{\lambda}_\text{batt}^\text{d}[k] +  \overline{\lambda}_\text{batt}^\text{d}[k] = \nu_\text{hes}[k]\,,~~\forall k \in\mathcal{K}\,,\label{eq:stat_discharge_simple}\\
    &-\underline{\lambda}_\text{batt}^\text{c}[k] +  \overline{\lambda}_\text{batt}^\text{c}[k] = \nu_\text{hes}[k]\,,~~\forall k \in\mathcal{K}\,.\label{eq:stat_charge_simple}
\end{align}

Next, we consider the case where the tracking error $e[k]:=Cr[k] - P_{\text{hes}}[k]$ is strictly positive at optimality for some time step~$k$ (i.e., $e[k]>0$).
From~\eqref{eq:stat_lam_w} and non-negativity of $\lambda$ dual variables, we have three possible cases: (i) $\underline{\lambda}_{w}[k] > 0$ and $\overline{\lambda}_{w}[k] > 0$; (ii) $\underline{\lambda}_{w}[k] = 1$ and $\overline{\lambda}_{w}[k] = 0$; or (iii) $\underline{\lambda}_{w}[k] = 0$ and $\overline{\lambda}_{w}[k] = 1$.
We will next show that cases (i) and (ii) lead to contradictions.
In case (i), the complementary slackness conditions associated with $\underline{\lambda}_{w}[k]$ and $\overline{\lambda}_{w}[k]$ force $w[k]=e[k]$ and $w[k]=-e[k]$.
This implies that $e[k]=0$, which contradicts our previous assumption ($e[k]>0$).
For case (ii), we have $w[k]=-e[k]$ (from complementary slackness for $\underline{\lambda}_{w}[k]$).
Substituting into~\eqref{eq:w_lims}, we have $e[k] \le -e[k]$, which is a contradiction if $e[k]>0$.
Therefore, we must have case (iii), where $\underline{\lambda}_{w}[k] = 0$ and $\overline{\lambda}_{w}[k] = 1$.
Moreover, from~\eqref{eq:nu_hes}, we have $\nu_\text{hes}[k]=1$.

Substituting $\nu_\text{hes}[k]=1$ into~\eqref{eq:stat_gen} and employing complementary slackness for $\underline{\lambda}_\text{gen}[k]$ and $\overline{\lambda}_\text{gen}[k]$, we have $\overline{\lambda}_\text{gen}[k] = 1$ and $P_\text{gen}[k]=\overline{P}_\text{gen}$.
Similary, from~\eqref{eq:stat_load},~\eqref{eq:stat_discharge_simple}, and~\eqref{eq:stat_charge_simple}, we have $P_\text{CL}[k]=0$, $P_\text{batt}^\text{d}[k]=\overline{P}_\text{batt}$, and $P_\text{batt}^\text{c}[k]=0$.
Thus, for $e[k]>0$, we have
\begin{equation}
\label{eq:error_pos}
    P_\text{hes}[k] = \overline{P}_\text{gen} + \overline{P}_\text{batt}~~\text{(if }e[k]>0\text{)}\,.
\end{equation}

Following a similar process, it can be shown that for $e[k]<0$, we have $\underline{\lambda}_{w}[k] = 1$, $\overline{\lambda}_{w}[k] = 0$, $\nu_\text{hes}[k]=-1$, $P_\text{gen}[k]=0$, $P_\text{CL}[k]=\overline{P}_\text{CL}$, $P_\text{batt}^\text{d}[k]=0$,  $P_\text{batt}^\text{c}[k]=-\overline{P}_\text{batt}$, and
\begin{equation}
\label{eq:error_neg}
    P_\text{hes}[k] = -\overline{P}_\text{CL} - \overline{P}_\text{batt}~~\text{(if }e[k]<0\text{)}\,.
\end{equation}
Combining~\eqref{eq:error_pos},~\eqref{eq:error_neg}, and the $e[k]=0$ case,
 the optimal HES dispatch from the convex problem (when SoC constraints are not binding) is
 \begin{equation}
 \label{eq:phes_opt}
 P_\text{hes}[k] = 
 \begin{cases}
     -\overline{P}_\text{CL} - \overline{P}_\text{batt}\,,& Cr[k] < -\overline{P}_\text{CL} - \overline{P}_\text{batt}\,,\\
     \overline{P}_\text{gen} + \overline{P}_\text{batt}\,,& Cr[k] > \overline{P}_\text{gen} + \overline{P}_\text{batt}\,,\\
     Cr[k]\,,& \text{otherwise}\,.\\
 \end{cases}
 \end{equation}
 Comparing~\eqref{eq:phes_opt} with Algorithm~\ref{alg:hes_dispatch_short}, we can verify that the output power of the HES of the real-time control is the same as the optimal HES output from the convex optimization problem when the SoC limits are not binding. 
 It is worth noting that while the total HES output (and therefore, the tracking error) is the same, the allocation of power to individual resources may be different between the online control strategy and the optimization problem.
 This is due to the fact that there are multiple (non-unique) ways to allocate power among the controllable generation, load, and battery to achieve $P_\text{hes}[k] = Cr[k]$.
 Since the HES power outputs are identical, we have
 \begin{equation}
    J_\text{on}(C,\mathbf{r}) = J^{*}_\text{cvx}(C,\mathbf{r})\,.
\end{equation}
Finally, note that the online control strategy respects the battery complementarity constraint~\eqref{eq:batt_cc} (i.e., it never simultaneously charges and discharges).
Therefore, under the assumption of Theorem~\ref{thm:online_optimality}, the online strategy gives an HES dispatch that is feasible for the original problem~\eqref{eq:offline_opt} and achieves the same objective function value as the convex relaxation~\eqref{eq:convex_prob}.  
Thus, $J_\text{on}(C,\mathbf{r}) = J^{*}_\text{cvx}(C,\mathbf{r}) = J^{*}_\text{off}(C,\mathbf{r})$.
This completes the proof.

\bibliographystyle{IEEEtran}
\bibliography{PSCC_Ref}
\end{document}